\documentclass[12pt]{article}
\usepackage[xdvi]{graphicx} 
\usepackage{epsfig}
\usepackage{amsfonts,amssymb,amsmath}

\newcommand{\al}{\ensuremath{\alpha}}

\newcommand{\la}{\ensuremath{\lambda}}
\newcommand{\La}{\ensuremath{\Lambda}}

\newcommand{\Om}{\ensuremath{\Omega}}

\newcommand{\s}{\ensuremath{\sigma}}

\renewcommand{\d}{\ensuremath{{\rm d}}}

\newcommand{\Del}{\ensuremath{\nabla}}

\newcommand{\be}{\begin{equation}}
\newcommand{\ee}{\end{equation}}

\newcommand{\ba}{\begin{eqnarray}}
\newcommand{\ea}{\end{eqnarray}}

\begin{document}

\rightline{hep-th/0304250}
\rightline{UUITP-06/03}
\rightline{OUTP-03/10P}
\vskip 1cm 

\begin{center}
{\Large \bf Braneworld holography in Gauss-Bonnet gravity}
\end{center}
\vskip 1cm
  
\renewcommand{\thefootnote}{\fnsymbol{footnote}}

\centerline{\bf
James P. Gregory$^\sharp$\footnote{James.Gregory@teorfys.uu.se}
  and 
Antonio Padilla$^\flat$\footnote{a.padilla1@physics.ox.ac.uk}}
\vskip .5cm

\centerline{\it $^\sharp$Department of Theoretical Physics}
\centerline{\it Uppsala University, Box 803, SE-751 08 Uppsala, Sweden}

\centerline{$\phantom{and}$}

\centerline{\it $^\flat$Theoretical Physics, Department of Physics}
\centerline{\it University of Oxford, 1 Keble Road, Oxford, OX1 3NP,  UK}

\setcounter{footnote}{0} \renewcommand{\thefootnote}{\arabic{footnote}}
 

\begin{abstract}
We investigate holography on an $(n-1)$-dimensional brane embedded in a
  background of AdS black holes, in $n$-dimensional Gauss-Bonnet
  gravity. We demonstrate that for a critical brane near the AdS
  boundary, the Friedmann equation corresponds to that of the standard
  cosmology driven by a CFT dual to the AdS bulk. We show that there
  is no holographic description for non-critical branes, or when the
  brane is further away from the AdS boundary. We then derive a
  Cardy-Verlinde formula for the dual CFT on the critical brane near
  the boundary. This gives us insight into the remarkable
  correspondence between Cardy-Verlinde formul{\ae} and Friedmann
  equations in Einstein gravity.
\end{abstract}

\newpage

\section{Introduction}

In recent years, there has been an enormous amount of research into
two important areas of theoretical physics: braneworld cosmology and
the holographic principle. The braneworld scenario gained momentum as
a solution to the hierarchy problem~\cite{Arkani:hier1,
Antoniadis:hier, Arkani:hier2, Randall:bw1}, although the single brane
model of Randall and Sundrum provided us with an interesting {\it
alternative to compactification}~\cite{Randall:bw2}. The holographic
principle, meanwhile, was first realised in string theory {\it via}
the AdS/CFT
correspondence~\cite{Maldacena:adscft,Witten:adscft,Gubser:adscft}.

At first glance, braneworld physics and holography are two very
distinct subjects. However, it was soon realised that this is not the
case~\cite{Verlinde:gravity,Gubser:gravity,Hawking:gravity}, and so
began the study of braneworld holography (see for
example~\cite{Hebecker:bwholog, Savonije:bwholog, Padilla:cft},
or~\cite{Padilla:thesis} for a review).

The essence of braneworld holography can be captured in the following
claim:  {\it Randall-Sundrum braneworld gravity is dual to a CFT with
a UV cutoff, coupled to gravity on the brane}.  Formal evidence for this
claim was provided by studying a brane universe in the background of
the Schwarzschild-AdS black hole.  The introduction of the black hole
on the gravity side of the AdS/CFT correspondence corresponds to
considering finite temperature states in the dual
CFT~\cite{Witten:thermal}.  In the context of braneworld holography,
Savonije and Verlinde demonstrated that their induced braneworld
cosmology could alternatively be described as the standard FRW
cosmology driven by the energy density of this dual
CFT~\cite{Verlinde:bwholog,Savonije:bwholog}.

In this article we develop this notion of braneworld holography to
include a broader class of bulk gravitational theories -- namely we
add the Gauss-Bonnet term to the standard Einstein-Hilbert action
giving
\be \label{eq:bulkaction}
S = \frac{1}{16\pi G_n}\int_\mathcal{M} \d^n x
\sqrt{-g}\lbrace R - 2\La_n + \al \mathcal{L}_{\mathrm{GB}}\rbrace,
\ee
where
\be
\mathcal{L}_{\mathrm{GB}} = R^2 - 4 R_{ab}R^{ab} + R_{abcd}R^{abcd}.
\ee
In $n=4$ dimensions, the Gauss-Bonnet term is a topological invariant
that does not enter the dynamics, but in $n=5$ or 6, the equations of
motion derived from this action include the Lovelock
tensor~\cite{Lovelock:lovelock}.  With the inclusion of this tensor,
these are then the most general equations of motion which satisfy the
same principles required for the construction of the Einstein-Hilbert
action in $n=4$.

It is therefore natural to want to consider the Gauss-Bonnet gravity
in higher dimensions from a purely classical point of view, for
generic values of the Gauss-Bonnet parameter, $\al$.  However, for
small values of $\al$, the study of Gauss-Bonnet gravity can be
motivated by string theory.  Curvature squared terms appear as the
leading order stringy correction to Einstein gravity in the $\al'$
expansion of the heterotic string action~\cite{Candelas:vac,
Green:Dinst}.  Furthermore, for this theory of gravity to be
ghost-free, the curvature squared terms must appear in the
Gauss-Bonnet combination~\cite{Zwiebach:gb, Zumino:theories,
Gross:heterotic}.  In the AdS/CFT correspondence, the introduction of
such higher order terms corresponds to next to leading order
corrections in the $1/N$ expansion of the
CFT~\cite{Fayyazuddin:ho,Aharony:ho, Nojiri:1/N}.  The importance of
Gauss-Bonnet gravity in the framework of braneworld holography is thus
self-evident. Previous studies of branes in higher derivative
theories of gravity suggest that no holographic description can be
found~\cite{Nojiri:gb1,Cho:gb2,Lidsey:gb1,Nojiri:gb2}. However,  we
demonstrate that a holographic description {\it does} exist in
Gauss-Bonnet gravity, at least for critical branes, close to the
boundary of AdS. This should come as no surprise. When the brane is
close to the AdS boundary the UV cutoff in the dual CFT is not too
significant, and one can justifiably appeal to the AdS/CFT
correspondence.

In Einstein gravity, we were able to relax the constraint on the
position of the brane~\cite{Gregory:exact}. By calculating the bulk
energy exactly {\it via} a Hamiltonian method, a larger equivalence
was observed.  When one reinterprets the black hole's contribution to
the braneworld cosmology as energy density due to the field theory,
nonlinear terms in the energy density and pressure are found in the
FRW equations.  These exactly reproduce those of the unconventional
cosmology~\cite{Binetruy:unconventional1,Binetruy:unconventional2}
described by a matterfilled brane in a pure AdS bulk.  We coin the
phrase ``{\it exact} holography'' to describe this equivalence.

The machinery to investigate exact holography in the Gauss-Bonnet
scenario has recently become available~\cite{Padilla:hamiltonian}.
Using this, we are able to show that in contrast to Einstein gravity,
a holographic description can {\it only} be found for flat branes near
the AdS boundary. This has two important implications regarding the
existence and behaviour of the Cardy-Verlinde formula for the dual
field theory. Firstly, in the limit of a valid holographic
description, it turns out that we can indeed cast the thermodynamic
properties of the dual CFT into a Cardy-Verlinde like formula,
provided we make consistent approximations. This is in contrast to
previous studies which suggest that no Cardy-Verlinde formula can be
found~\cite{Cai:cvgb,Cai:cv}. Having found this formula, we are able
to study its behaviour at the point that the brane crosses the black
hole horizon. In Einstein gravity, we find that we reproduce the
Friedmann equation! This does not happen in Gauss-Bonnet gravity. The
difference helps us to understand what is special about the Einstein
case. We believe that the remarkable correspondence between the
Friedmann equation and the Cardy-Verlinde formula in Einstein gravity
is related to the existence of exact holography.

The rest of this paper is organised as follows: In
section~\ref{sect:eom}, we describe the Gauss-Bonnet braneworld
scenario and review the derivation of the Friedmann equations in this
case.  In section~\ref{sect:ads}, we consider a brane moving in a pure
(Gauss-Bonnet) AdS bulk, with additional matter on the brane. This
enables us to establish the fine tuning condition for vanishing
braneworld cosmological constant, and to derive the connection between
the bulk and braneworld Newton constants. In
section~\ref{sect:holography}, we demonstrate that a holographic
description exists for a flat brane moving near the boundary of a
Gauss-Bonnet AdS black hole bulk. The cosmology is well described as
being the standard cosmology driven by a dual CFT. In
section~\ref{sect:exact}, we show that there is no exact holography in
Gauss-Bonnet gravity. In section~\ref{sect:CV}, we derive a
Cardy-Verlinde formula for the case where the holographic description
is valid. We comment that this doesn't make sense at the horizon, and
gain insight into the remarkable properties of the Cardy-Verlinde
formula in Einstein gravity.  Finally, section~\ref{sect:conc}
contains some concluding remarks.

\section{Equations of motion} \label{sect:eom}

Consider an $(n-1)$-dimensional brane moving in an $n$-dimensional
bulk, where $n \geq 5$. The bulk is a solution to Gauss-Bonnet gravity
with a negative (bare) cosmological constant, $\Lambda_n$. It is given
by two spacetimes, $\mathcal{M}_1$ and  $\mathcal{M}_2$, with
boundaries $\partial\mathcal{M}_1$ and $\partial\mathcal{M}_2$
respectively. The brane can be thought of as a domain wall between the
two spacetimes, so that it coincides with  $\partial\mathcal{M}_1$ and
$\partial\mathcal{M}_2$. For simplicity, we will assume that we have
$\mathbb{Z}_2$ symmetry across the brane. This scenario is described
by the following action,
\be \label{action}
S= S_\textrm{grav}+S_\textrm{brane},
\ee
where
\ba
S_\textrm{grav} &=&\frac{1}{16\pi G_n} 
  \int_{\mathcal{M}_1+\mathcal{M}_2} \d^n x
  \sqrt{-g}\lbrace R - 2\La_n + \al \mathcal{L}_{\mathrm{GB}}\rbrace
  \nonumber \\
  & & \qquad + \int_{\partial \mathcal{M}_1+ \partial \mathcal{M}_2}
  \textrm{boundary terms}, \\
S_\textrm{brane} &=& \int_{\textrm{brane}} d^{n-1}x
  \sqrt{-h}\, \mathcal{L}_\textrm{brane}.
\ea
The boundary integrals in $S_\textrm{grav}$ are required for a well
defined action principle~\cite{Myers:ghterm} (see
also~\cite{Davis:israel}). We denote the bulk  metric and the brane
metric by $g_{ab}$ and $h_{ab}$
respectively. $\mathcal{L}_\textrm{brane}$ describes the matter
content on the brane.

\subsection{The bulk}

For the action (\ref{action}), the bulk equations of motion are given
by
\be
R_{ab} - \frac{1}{2}R g_{ab} = -\La_n g_{ab} + \al
  \Big\{ \frac{1}{2}\mathcal{L}_{\mathrm{GB}}\,g_{ab}
    -2R R_{ab}+4R_{ac}{R_b}^c+4R_{acbd}R^{cd}-2R_{acde}{R_b}^{cde}
  \Big\}
\ee
Given the complexity of these equations, it is surprising that they
admit the following family of simple static black hole
solutions~\cite{Boulware:gbbh,Cai:gbbh} (see
also~\cite{Myers:HDBH,Charmousis:gb}):
\be \label{eq:gbbh}
\d s_n^2 =-h_\textrm{BH}(a) \d t^2 +\frac{\d a^2}{h_\textrm{BH}(a)}
  + a^2 \d \Om_{n-2}^2,
\ee
where $\d \Om_{n-2}^2$ is the metric on a unit $(n-2)$-sphere, and 
\be
h_\textrm{BH}(a)  =  1 + \frac{a^2}{2\tilde \al}\left(1 - \xi(a) \right)
  \quad \textrm{for} \quad \xi(a) =  \sqrt{1 - 4\tilde \al {k_n}^2 
  + \frac{4 \tilde \al\mu}{a^{n-1}}}.
\ee
By $\mathbb{Z}_2$ symmetry across the brane we have two identical
black holes, one each living on either side of the brane.  $\mu \geq
0$ is the constant of integration which determines the mass of each of
these black holes~\cite{Crisostomo:BHscan, Deser:quadenergy,
Deser:HDenergy, Padilla:hamiltonian},
\be
M=\frac{(n-2)\Omega_{n-2} \mu}{16 \pi G_n},
\ee
where $\Omega_{n-2}$ is the volume of a unit $(n-2)$-sphere. $k_n$ and
$\tilde \al$ are related to the bulk cosmological constant and the
Gauss-Bonnet parameter as follows,
\be \label{eq:conventions}
\La_n =  -\frac{1}{2}(n-1)(n-2){k_n}^2, \qquad \tilde \al =  (n-3)(n-4)\al
\ee

Since we could consider associating $\alpha$ with the slope parameter
({\it i.e.}  $\alpha'$) of heterotic string theory, from now on we
will assume that $\alpha \geq 0$. Furthermore, for the metric to be
real, we also have the condition
\be \label{eq:bound}
4\tilde \al {k_n}^2 \le 1.
\ee
These Gauss-Bonnet black hole solutions are asymptotically maximally
symmetric and, in the limit $\al \to 0$, they reduce to the standard
AdS black hole metric of Einstein gravity.

\subsection{The brane}

We now consider the dynamics of the brane, moving in the static black
hole bulk. The brane is given by the  section $(t(\tau), a(\tau), {\bf
x}^{\mu})$ of the bulk metric, where the parameter $\tau$ corresponds
to the proper time of an observer comoving with the brane. This gives
the condition
\be \label{unit}
-h_\textrm{BH}(a)\dot{t}^2 + \frac{\dot{a}^2}{h_\textrm{BH}(a)} = -1,
\ee
where overdot corresponds to differentiation with respect to
$\tau$. The induced metric is that of a FRW universe,
\be \label{eq:frw}
\d s_{n-1}^2 = -\d \tau^2 + a(\tau)^2 \d \Om_{n-2}^2,
\ee
with Hubble parameter $H=\dot a/a$. The equations of motion for the
brane are determined by the junction conditions for a braneworld in
Gauss-Bonnet gravity~\cite{Gravanis:israel,Davis:israel}. Given that
we have $\mathbb{Z}_2$ symmetry across the brane, these take the form
\be \label{eq:junction}
2(K_{ab} - K h_{ab}) + 4\al (Q_{ab} - \frac{1}{3} Q h_{ab}) 
   = - 8\pi G_n S_{ab},
\ee
where the energy momentum tensor on the brane is
\be
S_{ab}=-\frac{2}{\sqrt{h}}\frac{\delta S_\textrm{brane}}{\delta
  h^{ab}},
\ee
and 
\ba
Q_{ab} & = & 2K K_{ac}{K_b}^c -2K_{ac}K^{cd}K_{db} +
                K_{ab}(K_{cd}K^{cd}-K^2) \nonumber \\
& & \quad + 2K \mathcal{R}_{ab} + \mathcal{R} K_{ab}
  - 2 K^{cd}\mathcal{R}_{cadb} - 4\mathcal{R}_{ac}{K_b}^c.
\ea
For a brane with unit normal, $n_a$, the extrinsic curvature of the
brane is given by $K_{ab} = h_a^c h_b^d \Del_{\left(c\right. }
n_{\left. d \right)}$. $\mathcal{R}_{abcd}$ is the Riemann tensor on
the brane, constructed from the induced metric $h_{ab}$.

Since the brane is homogeneous and isotropic, its energy momentum
tensor is given in terms of its energy density, $\rho_\mathrm{brane}$,
and pressure, $p_\mathrm{brane}$, as follows,
\be
S_{ab} = (\rho_\mathrm{brane} + p_\mathrm{brane})\tau_a \tau_b +
p_\mathrm{brane} h_{ab},
\ee
where $\tau^a = (\dot{t}(\tau), \dot{a}(\tau), {\bf 0})$ is the
velocity of a comoving observer.  Given that the unit normal is $n_a =
(- \dot{a}(\tau), \dot{t}(\tau),{\bf 0})$, we can evaluate the
$\tau\tau$ component of~(\ref{eq:junction}) to give
\be \label{eq:dott}
\left(1 + \frac{4\tilde \al\dot{a}^2}{3 a^2} 
  + \frac{2\tilde \al}{a^2}\right) \frac{h_\textrm{BH}\dot{t}}{a}
  - \frac{2\tilde \al \left[h_\textrm{BH}\right]^2 \dot{t}}{3 a^3}
  =\frac{4 \pi G_n}{(n-2)}\rho_\mathrm{brane}.
\ee
If we square this equation, and use the condition~(\ref{unit}), we
obtain the following equation for the Hubble parameter\footnote{For
the original derivation of this equation, see~\cite{Charmousis:gb}.
Alternative forms of the Friedmann equation for a braneworld in the
Gauss-Bonnet bulk were derived
in~\cite{Kim:gbbw,Nojiri:gb1,Abdesselam:gbbw,Germani:gbbw}.}, $H$,
\be
\left[H^2 + \frac{h_\textrm{BH}}{a^2}\right]
  \left[ \frac{4\tilde \al}{3}H^2 + 1 +
  \frac{2\tilde \al}{a^2}\left(1-\frac{1}{3}h_\textrm{BH}\right)
  \right]^2 = 
  \left(\frac{4 \pi G_n}{n-2}\right)^2\rho_\mathrm{brane}^2.
\ee
This is a cubic equation for $H^2$, with one real solution. We can
extract this solution to write down the Friedmann equation for our
braneworld, in its standard explicit form.  To simplify the appearance
of the equation we make the following definitions,
\ba
\la & = & 3 \sqrt{\tilde \al}\left(\frac{4 \pi
  G_n}{n-2}\right)\,\rho_\mathrm{brane}, \\
\zeta_\pm (a) & = & \left( \sqrt{\la^2 + \xi(a)^3} \pm \la
\right)^{\frac{1}{3}}.
\ea
The Friedmann equation now reads
\be \label{eq:fried}
H^2 = - \frac{1}{a^2} + \frac{{\zeta_+}^2(a) + {\zeta_-}^2(a) -
  2}{4\tilde \al}.
\ee
If we expand this equation in $\tilde \al$, then to lowest (zeroth)
order, we recover the Friedmann equation for a brane in Einstein
gravity~\cite{Binetruy:unconventional1, Binetruy:unconventional2}, as
indeed we should, 
\be
H^2 = \left(\frac{4\pi G_n}{n-2}\right)^2\rho_\mathrm{brane}^2 - k_n^2
       - \frac{1}{a^2} +\frac{\mu}{a^{n-1}}.
\ee

\section{Brane moving in an AdS background} \label{sect:ads}

We now consider the case where the bulk spacetime is pure AdS
space. This corresponds to the case $\mu=0$ in our bulk
metric~(\ref{eq:gbbh}).  In other words, the bulk metric is now given
by
\be
\d s_n^2=-h_\textrm{AdS}(a)\d t^2+\frac{\d a^2}{h_\textrm{AdS}(a)}
  + a^2 \d \Omega_{n-2}^2,
\ee
where
\be
h_\textrm{AdS}(a) = 1 + k_\mathrm{eff}^2 a^2.
\ee
The effective cosmological constant is given by
$\La_\textrm{eff}=-\frac{1}{2}(n-1)(n-2)k_\textrm{eff}^2$. This
differs from the bare cosmological constant, $\La_n$, because of the
Gauss-Bonnet correction,
\be
k_\mathrm{eff}^2 = \frac{1-\beta}{2\tilde \al} \quad \mathrm{where}
 \quad \beta = \sqrt{ 1 - 4\tilde \al {k_n}^2}.
\ee
We now assume that the energy-momentum of the brane splits into a
contribution from the brane tension, $\sigma$, and a contribution from
additional matter fields. We write 
\be
\rho_\textrm{brane}=\rho+\sigma, \qquad p_\textrm{brane}=p-\sigma,
\ee
where $\rho$ and $p$ are the energy density and pressure respectively,
of the additional matter fields. If we define
\be
\tilde{\s} = 3 \sqrt{\tilde \al}\left(\frac{4\pi G_n}{n-2}\right) \s,
\ee
and 
\be
\zeta_\pm^\ast = \left( \sqrt{\tilde{\s}^2 + \beta^3} \pm \tilde{\s}
\right)^{\frac{1}{3}},
\ee
then we can expand the Friedmann equation (\ref{eq:fried}) about
$\rho=0$,
\be \label{eq:smallrho}
H^2 =  \mathcal{A} - \frac{1}{a^2} +
\frac{2\pi G_n({\zeta_+^\ast}^2 - {\zeta_-^\ast}^2)}
     {(n-2)\sqrt{\tilde \al}\sqrt{\tilde{\s}^2 + \beta^2}}
   \, \rho +\mathcal{O}(\rho^2), \\
\ee
where
\be
\mathcal{A}  =  \frac{{\zeta_+^\ast}^2 + {\zeta_-^\ast}^2 - 2}{4\tilde \al}.
\ee
For $\rho \ll \s$, this looks like the Friedmann equation for the
standard cosmology of an $(n-1)$-dimensional $\kappa=1$ universe,
\be \label{standard}
H^2= \mathcal{A} - \frac{1}{a^2}+\frac{16 \pi G_{n-1}}{(n-2)(n-3)}\rho,
\ee
where the cosmological constant $\La_{n-1}=\frac{1}{2}(n-2)(n-3)\mathcal{A}$. 

Let us restrict our attention to critical branes with vanishing
cosmological constant, $\mathcal{A}=0$. This requires us to fine tune
the brane tension in the following way,
\be \label{eq:fine}
\tilde{\s} = (2+\beta)\sqrt{\frac{1-\beta}{2}}.
\ee
In this case, the coefficient multiplying $\rho$ in the Friedmann
equation can be simplified dramatically, so that we now have
\be
H^2 = -\frac{1}{a^2} + \frac{8\pi G_n
  k_{\mathrm{eff}}}{(n-2)(2-\beta)}\,\rho +\mathcal{O}(\rho^2).
\ee
For $\rho \ll \s$, we can compare this with equation (\ref{standard})
to find an expression for the Newton's constant on the brane,
\be
G_{n-1} =  \frac{(n-3)G_n k_\mathrm{eff}}{2(2-\beta)}.
\label{eq:nc}
\ee
For the five dimensional bulk, this agrees with the expression derived
in~\cite{Cho:gb1} (see also~\cite{Neupane:newton}).  Furthermore, for
the case of general $n$, it agrees with the standard relation for
critical branes in the $\al\to 0$ limit~\cite{Giddings:linear}.

\section{Braneworld holography} \label{sect:holography}

We will now consider the case where $\mu>0$, that is, when the brane
is moving in a Gauss-Bonnet black hole bulk. We will assume that there
is no additional matter on the brane, so that its energy-momentum
only contains brane tension,
\be
\rho_\textrm{brane}=\sigma, \qquad p_\textrm{brane}=-\s.
\ee
For Einstein gravity ($\al=0$), it has been shown that when the brane
is near the AdS boundary, we can think of its dynamics as being
described by a radiation dominated FRW universe. This radiation is
given by a strongly coupled CFT with an AdS dual
description~\cite{Savonije:bwholog}.

It is natural to ask if these ideas can be extended to branes moving
in a Gauss-Bonnet bulk, with $\al \sim k_n^{-2}$. In this section, we
will demonstrate that, for critical branes ($\mathcal{A}=0$) near the AdS
boundary, they can.

We begin by clarifying what we mean by ``near the AdS boundary''. We
mean that the brane position is given by $a(\tau) \gg a_H$, where
$a_H$ is the radius of the black hole horizon
($h_\textrm{BH}(a_H)=0$). However, for critical branes
($\mathcal{A}=0$) and anti-de Sitter branes ($\mathcal{A}<0$), the
trajectory will have a maximum value of $a$. In order to have $a \gg
a_H$, we require that $a_H \gg \sqrt{\al}$. In fact, a discussion of
anti-de Sitter branes  cannot be included as the large $a$ limit also
requires $|\mathcal{A}| \ll \al^{-1}$ when $\mathcal{A}<0$ (see
appendix~\ref{app}).

For $a \gg a_H$, the Friedmann equation (\ref{eq:fried}) can be
approximated\footnote{We are assuming $1>4 \tilde \al k_n^2$, although
the argument presented in this section {\it can} be modified to
include $1=4 \tilde \al k_n^2$.} by
\be \label{jamesisgay}
H^2 = \mathcal{A}- \frac{1}{a^2} + 
  \left[\frac{\zeta_+^\ast + \zeta_-^\ast}{2\sqrt{{\tilde{\s}}^2+\beta^3}}
  \right]\frac{\mu}{a^{n-1}}.
\ee
As in section~\ref{sect:ads}, we restrict attention to critical
  branes.  The fine-tuning of the brane tension~(\ref{eq:fine})
  simplifies the coefficient of $\mu$, so that our Friedmann equation
  becomes
\be \label{eq:frwcrit}
H^2 = - \frac{1}{a^2} + \frac{\mu}{(2-\beta)a^{n-1}}.
\ee
Our main interest lies in the contribution from the black hole
  masses. As in Einstein gravity, we will show that this contribution
  can be thought of as coming from the energy density of a dual CFT --
  we will now calculate this energy density.

The energy of the bulk is given by the sum of the black hole masses,
$E_\textrm{bulk}=2M$.  This energy is measured with respect to the
bulk time coordinate, $t$, whereas an observer on the brane measures
energy with respect to the brane time coordinate, $\tau$. To arrive at
the energy of the CFT, we therefore need to scale the bulk energy by
$\dot t$, $E_\textrm{CFT}=E_\textrm{bulk} \dot t$.

This redshift factor can be found using equation (\ref{eq:dott}). Given
that we are near the AdS boundary we find that
\be
\dot t \approx \frac{2\sqrt{\tilde \al}\tilde \s}{(1-\beta)(2+\beta)a}.
\ee
We now impose the fine tuning condition (\ref{eq:fine}), to give
\be
\dot t \approx \frac{1}{k_\mathrm{eff}a}.
\ee
The CFT energy is therefore given by
\be
E_{CFT} = 2M\dot{t} \approx \frac{(n-2)\Om_{n-2}\mu}
{8\pi G_n k_\mathrm{eff}a}.
\ee
To calculate the energy density, we need to divide by the spatial
volume of the CFT,
\be
V_{CFT} =\Om_{n-2}a^{n-2}.
\ee
Finally we arrive at the following expression for the energy density
of the dual CFT,
\be \label{eq:rho}
\rho_{CFT} = \frac{(n-2)\mu} {8\pi G_n k_\mathrm{eff}a^{n-1}}.
\ee
We now rewrite the Friedmann equation (\ref{eq:frwcrit}) in terms of
this energy density.
\be \label{eq:frwstand}
H^2 =  - \frac{1}{a^2} +
   \frac{8\pi G_n k_\mathrm{eff}}{(n-2)(2-\beta)} \, \rho_{CFT}.
\ee  
Using the relation between the braneworld and bulk Newton's
constants~(\ref{eq:nc}), the Friedmann equation becomes,
\be
H^2 =  - \frac{1}{a^2} + 
   \frac{16\pi G_{n-1}}{(n-2)(n-3)}\,\rho_{CFT}.
\ee
This is just the Friedmann equation for the standard cosmology in
$(n-1)$ dimensions. The cosmology is driven by a strongly coupled CFT,
which is dual to the AdS black hole bulk. For critical branes near the
AdS boundary, we conclude that there is a holographic description even
when the bulk gravity includes a Gauss-Bonnet correction.

We should also note that we can use the thermodynamic relation,
$p=-\partial E/\partial V$, to derive the CFT pressure from the energy
density. The equation of state corresponds to that of radiation,
\be \label{pressure}
p_\textrm{CFT}=\frac{\rho_\textrm{CFT}}{n-2}.
\ee
If we differentiate the Friedmann
equation~(\ref{eq:frwcrit}) with respect to $\tau$, then the resulting
equation,
\be
\dot{H} = \frac{1}{a^2} - \frac{(n-1)\mu}{2(2-\beta)a^{n-1}},
\ee
can be written as the second of the FRW equations of the standard
cosmology,
\be
\dot{H} = \frac{1}{a^2} - \frac{8 \pi G_{n-1}}{(n-3)}\left
(\rho_{CFT}+p_{CFT}\right).
\ee

\section{Exact holography?} \label{sect:exact}

The AdS/CFT correspondence relates gravity on $n$-dimensional AdS
space to a CFT on $(n-1)$-dimensional Minkowski space. In braneworld
holography, the field theory on the brane is cutoff in the UV. This
cutoff vanishes as the brane approaches the boundary of AdS so that
the field theory becomes conformal. Only at this point can we
confidently appeal to the AdS/CFT correspondence. Although it is
natural to expect a holographic description for critical branes near
the AdS boundary there is no reason to expect more. However, in
Einstein gravity, it has been shown that a holographic description
exists for non-critical branes~\cite{Padilla:cft}. Perhaps even more
surprisingly, there is a form of {\it exact} holography in Einstein
gravity~\cite{Gregory:exact}. This is where the condition that the
brane should be near the AdS boundary is relaxed. In this section we
ask whether or not the same generalisations can be made in
Gauss-Bonnet gravity.

\subsection{Exact holography in Einstein gravity}

We start by reviewing precisely what we mean by exact holography
in Einstein gravity. Consider a brane moving in pure AdS space, with
\be
h_\textrm{AdS}(a)=k_n^2a^2+1.
\ee
As in section~\ref{sect:ads}, we assume that the energy-momentum of
the brane is made up of tension, $\s$, and additional matter with
energy density, $\rho$, and pressure, $p$. The Friedmann equation
is~\cite{Binetruy:unconventional1, Binetruy:unconventional2}
\be \label{eq:frwexact1}
H^2=\mathcal{A}-\frac{1}{a^2}+\frac{16 \pi G_{n-1}}{(n-2)(n-3)}\rho\left[1 +
\frac{\rho}{2\s}\right].
\ee
This takes the form of the $(n-1)$-dimensional standard cosmology when
$\rho \ll \s$.

Now consider a brane with no additional matter, moving in an AdS black
hole bulk, with
\be
h_\textrm{BH}(a)=k_n^2 a^2+1-\frac{\mu}{a^{n-3}}.
\ee
In this case the Friedmann equation is given by
\be \label{eq:frwexact2}
H^2=\mathcal{A}-\frac{1}{a^2}+\frac{\mu}{a^{n-1}}.
\ee
In~\cite{Gregory:exact}, we showed how we can calculate {\it exactly}
the energy density, $\rho_\textrm{FT}$, measured by an observer on the
brane -- this can be done {\it without} assuming that the brane is
near the AdS boundary. $\rho_\textrm{FT}$ is given in terms of $\mu$,
so we can rewrite the Friedmann equation (\ref{eq:frwexact2}) to give
\be
H^2=\mathcal{A}-\frac{1}{a^2}+\frac{16 \pi G_{n-1}}{(n-2)(n-3)}
 \rho_\textrm{FT}\left[1 + \frac{\rho_\textrm{FT}}{2\s}\right].
\ee
This takes exactly the same form as the Friedmann
  equation~(\ref{eq:frwexact1}) for the brane moving in pure AdS space
  with additional matter on the brane. We can therefore think of
  $\rho_\textrm{FT}$ as being the energy density of a field theory
  living on the brane. This field theory is dual to the AdS black hole
  bulk, although it is no longer conformal.  We think of the dual
  field theory on the brane as being cut off in the ultra violet -- this
  cutoff disappears as we go closer and closer to the AdS boundary,
  and we approach a conformal field theory.  In this case, we are not
  assuming that the brane is near the boundary, so the cutoff can be
  significant.

\subsection{Exact holography in Gauss-Bonnet gravity?}

We shall now investigate whether or not exact holography exists when
the bulk is a Gauss-Bonnet black hole. Our result relies on the
expression for the Gauss-Bonnet Hamiltonian found
in~\cite{Padilla:hamiltonian}. We will find it convenient to rederive
the Friedmann equation from the action given in that paper.

Consider the timelike vector field defined on the brane,
\be \label{tau}
\tau^a=(\dot t,  \dot a, {\bf 0}).
\ee
This maps the brane onto itself and satisfies $\tau^a \nabla_a
\tau=1$. In principle we can extend the definition of $\tau$ into the
bulk, stating only that it approaches the form given by equation
(\ref{tau}) as it nears the brane. Now introduce a family of spacelike
surfaces, $\{ \Sigma_\tau \}$, each labelled by the parameter $\tau$,
so that we have a foliation of the bulk spacetime. These surfaces
should meet the boundary/brane orthogonally. We decompose $\tau^a$
into the lapse function and shift vector,
\begin{equation}
\tau^a=N r^a +N^a,
\end{equation}
where $r^a$ is the unit normal to  $\Sigma_\tau$.

We can now write the bulk metric in ADM form
\begin{equation}
ds_n^2=g_{ab}dx^adx^b=-N^2 d\tau^2
   +\gamma_{ab}(dx^a+N^a d\tau)(dx^b+N^b d\tau),
\end{equation}
where $\gamma_{ab}$ is the induced metric on  $\Sigma_\tau$.

If $S_\tau$ is the intersection of the brane and $\Sigma_\tau$, then
the family of surfaces $\{ S_\tau \}$ is a foliation of the brane. In
the same way as for the bulk, we can write the brane metric in ADM
form,
\begin{equation}
ds^2_{n-1}=h_{ab}dx^a dx^b=-N^2d\tau^2
  +\lambda_{ab}(dx^a+N^a d\tau)(dx^b+N^b d\tau),
\end{equation}
where $\lambda_{ab}$ is the induced metric on $S_\tau$.

Now, from~\cite{Padilla:hamiltonian}, we can write the action as
\begin{equation}
S=S_\textrm{grav}+S_\textrm{brane},
\end{equation}
in which
\begin{equation}
S_\textrm{grav}=\frac{1}{8\pi G_n}\int dt \left\{\int_{\Sigma_t} 
  d^{n-1}x\left(\pi^{ab} \dot \gamma_{ab} - N\mathcal{H}
  - N^a \mathcal{H}_a\right)-\int_{S_t} d^{n-2}x\sqrt{\lambda}
  \left(N\mathcal{J}+N^a\mathcal{J}_a \right)\right\},
\end{equation}
where  $\pi^{ab}$ is the momentum conjugate to $\gamma_{ab}$, and
$\mathcal{H}$ and $\mathcal{H}_a$ are the Hamiltonian and momentum
constraints respectively. We have an overall factor of $2$ by
$\mathbb{Z}_2$ symmetry across the brane.

The $\tau \tau$ component of the junction conditions at the brane
gives rise to the Friedmann equation. This is obtained by varying the
brane part of the action with respect to the brane metric. For the
$\tau \tau$ component this amounts to variation with respect to $N$,
which gives

\begin{equation}
J + N\left(\frac{\delta \mathcal{J}}{\delta N}\right)
  +N^a \left(\frac{\delta \mathcal{J}_a}{\delta N}\right)
  =\frac{8 \pi G_n}{\sqrt{\lambda}}\frac{\delta S_\textrm{brane}}
  {\delta N}.
\end{equation}
The energy-momentum tensor on the brane is given by
\begin{equation}
S^{ab}=\frac{2}{\sqrt{h}} \frac{\delta S_\textrm{brane}}{\delta h_{ab}}.
\end{equation}
Furthermore, since $h_{\tau\tau}=-N^2$ and $\sqrt{h}=N\sqrt{\lambda}$,
we find that
\begin{equation}
\frac{\delta S_\textrm{brane}}{\delta N}=-N^2\sqrt{\lambda}S^{\tau\tau},
\end{equation}
so that the Friedmann equation now reads
\begin{equation}
J + N\left(\frac{\delta \mathcal{J}}{\delta N}\right)
  + N^a \left(\frac{\delta \mathcal{J}_a}{\delta N}\right)
  = -8\pi G_nN^2S^{\tau\tau}.
\end{equation}
Note that we have a homogeneous isotropic brane,
\begin{equation}
ds^2_{n-1}=-d\tau^2 +a^2 d \Omega_{n-2}^2,
\end{equation}
moving in a static bulk,
\begin{equation} \label{BHmetric}
ds^2_n=-h(a)dt^2+\frac{da^2}{h(a)}+a^2 d \Omega_{n-2}^2.
\end{equation}
Now, on the brane, $\tau^a$ is in fact the unit normal to $S_t$. This
means that $N=1$ and $N^a=0$ on the brane, although this need not be
the case in the bulk. We also have $S^{\tau
\tau}=\rho_\textrm{brane}$. Putting all this information back into the
Friedmann equation, we now have
\begin{equation}
\mathcal{J} + \mathcal{J}'=-8\pi G_n\rho_\textrm{brane},
  \qquad \textrm{where} \qquad
  \mathcal{J}'=\frac{\delta \mathcal{J}}{\delta N}.
\end{equation}

Now consider two different scenarios: ({\it i}) a bulk AdS black hole,
with the brane matter made up of tension only, and ({\it ii}) a pure
AdS bulk, with the brane matter made up of tension, and some
additional matter such as radiation.  For case ({\it i}), we have
$h(a)=h_\textrm{BH}(a)$ and $\rho_\textrm{brane}=\sigma$, where
$\sigma$ is the brane tension.. For case ({\it ii}), we  have
$h(a)=h_\textrm{AdS}(a)$ with $\rho_\textrm{brane}=\sigma+\rho$, where
$\rho$ is the energy density of the additional matter. From a
holographic point of view, we would expect these two cases to be
equivalent if the energy density, $\rho$, corresponds to that of a
field theory dual to the bulk black hole gravity.

Let us begin by considering the case with the bulk black hole. The
Friedmann equation reads
\begin{equation} \label{BHFRW}
\mathcal{J}_\textrm{BH} + \mathcal{J}'_\textrm{BH}=-8\pi G_n\sigma.
\end{equation}
In order to see if we have a holographic description in the way we
have just described, we need to calculate the energy density of the
bulk, as measured by an observer on the brane -- in other words, we want
to calculate the energy density of the bulk using $\tau$ as our time
coordinate. This is done by evaluating the Hamiltonian with a suitable
choice of background. We choose the background, $\mathcal{\bar M}$,
to be pure AdS space (with the same effective cosmological constant as
the black hole spacetime),
\begin{equation}
ds^2_\textrm{AdS}=-h_\textrm{AdS}(a)
dT^2+\frac{da^2}{h_\textrm{AdS}(a)}+a^2 d\Omega_{n-2}^2,
\end{equation}
cut off at a surface, $\partial\mathcal{\bar M}$,  given by 
\begin{equation}
T=T(\tau), \quad a=a(\tau) \qquad \textrm{where} \quad
-h_\textrm{AdS}(a)\dot T^2+\frac{\dot a^2}{h_\textrm{AdS}(a)}=-1.
\end{equation}
This ensures that the geometry on $\partial\mathcal{\bar M}$ is the
same as that on the brane.

Given that $N=1$ and $N^a=0$ on the brane, we evaluate the Hamiltonian
to derive the energy of the bulk measured with respect to $\tau$,
\begin{equation}
\mathcal{E}=\frac{1}{8\pi G_n} \int_{S_\tau} d^{n-2}x \sqrt{\lambda}
\left(\mathcal{J}_\textrm{BH}-\mathcal{J}_\textrm{AdS} \right),
\end{equation}
where we have a factor of two because there are two copies of the
bulk. To get the energy density we need to divide by the spatial
volume of the brane,
\begin{equation}
V=\int_{S_\tau} d^{n-2}x \sqrt{\lambda}.
\end{equation}
We thus see that
\begin{equation}
\rho=\frac{\mathcal{E}}{V}=\frac{1}{8\pi G_n}
  (\mathcal{J}_\textrm{BH}-\mathcal{J}_\textrm{AdS}).
\end{equation}
If we substitute this back into the Friedmann equation (\ref{BHFRW}),
we obtain
\begin{equation}
\mathcal{J}_\textrm{AdS}+\mathcal{J}'_\textrm{AdS}
  +(\mathcal{J}'_\textrm{BH}-\mathcal{J}'_\textrm{AdS})
  =-8\pi G_n(\sigma+\rho).
\end{equation}
It is clear that for the holographic description to be valid, we need
to ignore the contribution from
$(\mathcal{J}'_\textrm{BH}-\mathcal{J}'_\textrm{AdS})$. This is
achieved if we satisfy the condition
\begin{equation} \label{condition}
|\mathcal{J}| \gg |\mathcal{J}'|.
\end{equation}

For Gauss-Bonnet gravity, we have~\cite{Padilla:hamiltonian}
\begin{equation} \label{J}
\mathcal{J} =2K+12  \alpha \delta^{[l}_a\delta^{m}_b\delta^{n]}_c
 K^a_l \left[ \widehat R^{bc}{}_{mn}-2H^b_mH^c_n 
-\frac{2}{3} K^b_mK^c_n \right],
\end{equation}
where $ \widehat R^{bc}{}_{mn}$ is the Riemann tensor on $S_\tau$,
$K^a_b$ is the extrinsic curvature of $S_\tau$ in $\Sigma_\tau$ and
$H^a_b$ is the extrinsic curvature of $S_\tau$ in the brane.  To
evaluate $\mathcal{J}'$, we need to vary equation (\ref{J}) with
respect to $N$. However, the only term that depends on $N$ is $H^a_b$,
which is proportional to $1/N$.  Therefore,
\begin{equation}
\mathcal{J}'=\frac{48}{N}  \alpha
\delta^{[l}_a\delta^{m}_b\delta^{n]}_c K^a_lH^b_mH^c_n.
\end{equation}
We see immediately that when $\alpha=0$, then $\mathcal{J}'=0$, and
the condition (\ref{condition}) holds for all values of
$a(\tau)$. This is why we have an exact holographic description for
Einstein gravity. This is surprising because we would expect the
AdS/CFT description to only be valid when the brane is near the AdS
boundary where the cutoff to the dual CFT is insignificant.

What happens when we consider Gauss-Bonnet gravity explicitly, and
$\alpha \sim k_n^{-2}$~? To evaluate $\mathcal{J}$ and $\mathcal{J}'$,
we need the following, 
\begin{equation} \label{stuff}
\widehat
R^{bc}{}_{mn}=\frac{1}{a^2}
  \left(\delta^b_m\delta^c_n-\delta^c_m\delta^b_n\right),
  \qquad H_{ab}=-H \lambda_{ab},
  \qquad K_{ab}=\sqrt{\frac{h(a)}{a^2}+H^2}
  ~\lambda_{ab},
\end{equation}
where $H$ is the Hubble parameter, and $h(a)$ is $h_\textrm{BH}(a)$ or
$h_\textrm{AdS}(a)$, depending on whether we are working with the BH  or
AdS spacetime. We shall leave $h(a)$ general in what follows.

Given the equations stated in (\ref{stuff}), we find that
\begin{eqnarray}
\mathcal{J} &=& 2(n-2)\sqrt{H^2+\frac{h(a)}{a^2}}\left\{1+2\tilde \alpha
\left[\frac{1}{a^2}-\frac{4}{3}H^2-\frac{1}{3}\frac{h(a)}{a^2}
\right]\right\}, \\
\mathcal{J}' &=& 8(n-2)\tilde \alpha \sqrt{H^2+\frac{h(a)}{a^2}}~H^2,
\end{eqnarray}
where $\tilde \alpha = (n-3)(n-4)\alpha$, as adopted
in~(\ref{eq:conventions}).  For $\alpha \sim k_n^{-2}$, the condition
(\ref{condition}) amounts to,
\begin{equation} \label{cond4}
\al^{-1} \gg H^2.
\end{equation}
We are now ready to ask if we can have {\it exact} holography like we
did for Einstein gravity. For $a \sim \sqrt{\al}$, it is clear that
both sides of (\ref{cond4}) are of order $\al^{-1}$, and the condition
(\ref{condition}) does not hold. There is no {\it exact} holography in
Gauss-Bonnet gravity. This is not really surprising -- the fact that
we found exact holography for Einstein gravity was remarkable. As we
suggested earlier, {\it via} the AdS/CFT motivation for braneworld
holography, we would only really expect to find a holographic
description for critical branes near the AdS boundary.

Given that we know that the condition (\ref{condition}) doesn't hold
for general values of $a$, we now ask what happens when $a$ is
large. Recall that we must immediately eliminate the possibility of
anti-de Sitter branes ($\mathcal{A}<0$). For critical branes and de
Sitter branes, $H^2 \sim \mathcal{A}$ for large $a$, so the condition
(\ref{condition}) only holds if we have
\begin{equation}
\al^{-1} \gg \mathcal{A}.
\end{equation}
This suggests that we would only find a holographic description for
critical branes satisfying $\mathcal{A}=0$. Unlike in Einstein
gravity, there will be no extension to non-critical branes.

To sum up, we have found a condition that determines when a
holographic description will be valid for a brane moving in a black
hole bulk. This condition is satisfied for Einstein gravity,
regardless of the brane's position, or the value of the braneworld
cosmological constant. For Gauss-Bonnet gravity, the condition is far
more restrictive. It only holds for critical branes close to the AdS
boundary. We conclude that the holographic description shown in the
last section is the only one you can find for branes in  Gauss-Bonnet
gravity.

\section{Cardy-Verlinde formul{\ae}} \label{sect:CV}

The Cardy-Verlinde formula~\cite{Verlinde:bwholog} for a CFT is the
generalisation to arbitrary dimensions of the well known Cardy
formula~\cite{Cardy:cardy} for $1+1$-dimensional CFTs. It relates the
entropy, $S$, of the CFT, to its energy, $E$, and Casimir energy,
$E_c$. The Casimir energy can be thought of as providing  the
non-extensive part of the formula. In this paper, we will discuss the
local version of the Cardy-Verlinde formula for a CFT living on an FRW
brane.

We begin with the thermodynamic relation
\be
\d E= T \,\d S-p \,\d V,
\ee
where $T$, $p$ and $V$ are the temperature, pressure and volume of the
CFT respectively.
If we introduce the following densities,
\be
s=\frac{S}{V}, \qquad \rho=\frac{E}{V},
\ee
we can use the fact that $V \sim a^{n-2}$ to rewrite the thermodynamic
relation in the following form,
\be
d \rho=T ds+\gamma d\left(\frac{1}{a^2}\right),
\ee
where
\be
\gamma=\frac{(n-2)a^2}{2}(\rho + p-Ts).
\ee
We can think of $\gamma$ as describing the variation of $\rho$ with
respect to the spatial curvature, $1/a^2$. If the entropy and energy
were purely extensive, $\gamma$ would vanish. $\gamma$ will therefore
give the non-extensive part of the local Cardy-Verlinde formula.

In section \ref{sect:ads}, we showed that for a critical brane near
the AdS boundary, the dynamics is driven by a CFT that is dual to the
Gauss-Bonnet AdS black hole bulk. Eventually we will state the
Cardy-Verlinde formula for this CFT. However, first we will review the
form of the Cardy-Verlinde formula for a CFT that is dual to an AdS
black hole bulk in Einstein gravity.
 
\subsection{Cardy-Verlinde formul{\ae} for CFTs with AdS duals in
  \protect{\newline}Einstein gravity}

Now consider a critical brane moving in a black hole bulk, in Einstein
gravity\cite{Savonije:bwholog}. When it is near the AdS boundary, the
CFT on the brane obeys the following Cardy-Verlinde formula,
\be \label{CVEH}
s^2=\left(\frac{4 \pi}{n-2}\right)^2\gamma \left( \rho -
\frac{\gamma}{a^2} \right).
\ee
A remarkable connection between this formula, and the Friedmann
equation was noted in~\cite{Savonije:bwholog}. At the point that the
brane crosses the horizon, the entropy density is given by the Hubble
entropy,
\be
s=\frac{(n-3)H}{4 G_{n-1}},
\ee
and $\gamma=(n-2)(n-3)/16\pi G_{n-1}$. The Cardy-Verlinde formula now
reads
\be
H^2=-\frac{1}{a^2}+\frac{16 \pi G_{n-1}}{(n-2)(n-3)}\rho.
\ee
This is precisely the Friedmann equation for the standard cosmology!

However, we should be cautious. Contrary to the claim made in
\cite{Savonije:bwholog}, the formula (\ref{CVEH}) is only valid when
$k_n^{-1} \ll a_H \ll a$. This must be the case because the dual field
theory ceases to be conformal for smaller values of $a$. We would
therefore expect the structure of the formula to change, and an extra
scale to be introduced, reflecting the fact that conformal invariance
has been broken.

Since the holographic description exists in Einstein gravity for all
values of $a$, it was possible to find a more exact version of
this formula~\cite{Gregory:exact},
\be \label{exactCV}
s^2=\left( \frac{4 \pi}{n-2} \right)^2\gamma
  \left(1+\frac{\rho}{\s}\right)
  \left[ \rho \left(1+\frac{\rho}{2\s}
  \right)-\frac{\gamma}{a^2}\left(1+\frac{\rho}{\s} \right) \right].
\ee
The brane tension, $\s$ now appears as the new scale in the
formula. For $\rho \ll \s$, this formula reduces to its conformal
version (\ref{CVEH}).

It is interesting to evaluate this formula at the point that the brane
crosses the horizon. Once again, the entropy density is given by the
Hubble entropy, but this time we have a more precise formula for
$\gamma$. It is given by~\cite{Gregory:exact}
\be
\gamma \left(1+\frac{\rho}{\s}\right)=\frac{(n-2)(n-3)}{16 \pi
  G_{n-1}}.
\ee
The generalised Cardy-Verlinde formula (\ref{exactCV}) now coincides
with the exact braneworld Friedmann equation (\ref{eq:frwexact1}).

\subsection{Cardy-Verlinde formul{\ae} for CFTs with AdS duals in
  \protect{\newline}Gauss-Bonnet gravity}

Now consider the case of the critical brane moving in a Gauss-Bonnet
black hole bulk. For the brane near the AdS boundary we have a
holographic description, but not otherwise. We shall now derive the
Cardy-Verlinde formula for the CFT on the brane, when the holographic
description holds.

The entropy of the CFT is just given by the total entropy of the two
Gauss-Bonnet black holes~\cite{Cai:gbbh}~(see
also~\cite{Neupane:entropy1,Neupane:entropy2}),
\be \label{GBentropy}
S=2.\frac{\Omega_{n-2}a_H^{n-2}}{4 G_n} \left[1+2
\left(\frac{n-2}{n-4}\right)\frac{\tilde \al}{a_H^2} \right].
\ee
Note that the Gauss-Bonnet black hole entropy does not obey the area
law which exists in Einstein gravity. The temperature of the CFT is
given by the temperature of the black hole, $T_\textrm{BH}$, with the
appropriate redshift factor, $\dot t \approx 1/k_\textrm{eff}a$,
\be
T=\frac{T_\textrm{BH}}{k_\textrm{eff}a}, \quad \textrm{where} \quad
T_\textrm{BH}=\frac{h_\textrm{BH}{}'(a_H)}{4\pi}.
\ee
In section \ref{sect:holography}, we found the energy density and
pressure of the CFT (equations (\ref{eq:rho}) and (\ref{pressure})
respectively). In principle we can now calculate $\gamma$, and attempt
to construct a Cardy-Verlinde formula in the form of equation
(\ref{CVEH}). However, as noted in~\cite{Cai:cv} this will be impossible
if one attempts to include all the Gauss-Bonnet corrections.

Before we lose hope, it is important to take stock of what we are
actually trying to do. We are trying to find a Cardy-Verlinde formula
for a CFT that is dual to a Gauss-Bonnet AdS bulk. It only makes sense
to think of this CFT when the holographic description is valid, that
is when $\sqrt{\al} \ll a_H \ll a$. It is therefore inappropriate to
include the $\tilde \al/a_H^2$ correction in the entropy formula
(\ref{GBentropy}). In the holographic limit we can make the following
consistent approximations:
\ba
s &=& \frac{1}{2G_n} \left(\frac{a_H}{a}\right)^{n-2} \left[ 1+
\mathcal{O} \left(\frac{\tilde \al}{a_H^2}\right) \right] \\
\rho &=& \frac{n-2}{8 \pi G_n \tilde \al  k_\textrm{eff}}
\left(\frac{a_H}{a}\right)^{n-1} \left[ \tilde \al k_n^2 +\mathcal{O}
\left(\frac{\tilde \al}{a_H^2}\right) \right] \\
\gamma &=&  \frac{n-2}{8 \pi G_n  
k_\textrm{eff}}\left(\frac{a_H}{a}\right)^{n-3}\left[ \chi +\mathcal{O}
\left(\frac{\tilde \al}{a_H^2}\right) \right]
\ea
where  $\chi=1-2\left(\frac{n-1}{n-4}\right)\tilde \al k_n^2$. We can
now cast these quantities into a Cardy-Verlinde formula:
\be \label{CVGB}
s^2 = \frac{1}{\chi} \left( \frac{k_\textrm{eff}}{k_n}\right)^2
  \left(\frac{4\pi}{n-2}\right)^2
  \gamma\left[\rho-\frac{\gamma}{a^2}\right]\left(1+\mathcal{O}
\left(\frac{\tilde \al}{a_H^2}\right)\right),
\ee
Note that this formula agrees with~(\ref{CVEH}) in the limit $\al \to
0$.

We can now ask whether this formula bears any resemblance to the
Friedmann equation at the point  that the brane crosses the black hole
horizon -- the answer is no. However, it doesn't make sense to evaluate
this formula at $a=a_H$, as it is only valid for $a \gg a_H$. In
Einstein gravity, this was also the case, but the Cardy-Verlinde
formula (\ref{CVEH}) evaluated at the horizon, still gave the Friedmann
equation of the standard cosmology. We believe we now understand why
this was the case.

From a holographic perspective, we need to ask what is the difference
between an Einstein bulk and a Gauss-Bonnet bulk. The difference lies
in the existence of exact holography for Einstein gravity but not for
Gauss-Bonnet gravity. This means we cannot say anything sensible about
the CFT at the time the brane crosses the horizon for the Gauss-Bonnet
bulk. For the Einstein bulk, we can happily trust the exact
holographic description and use the generalised version of the
Cardy-Verlinde formula (\ref{exactCV}) at $a=a_H$. This formula agrees
with the braneworld Friedmann equation at each order of $\rho$. Since
the braneworld Friedmann equation and the standard Friedmann equation
agree up to order $\rho$, it is clear what is happening when we
evaluate the approximate formula (\ref{CVEH}) at the horizon. We are
just seeing  the agreement of (\ref{exactCV}) with the braneworld
Friedmann equation, up to order $\rho$. However, we should note that
the $\rho^2$ corrections are {\it not} small at the horizon.

In Gauss-Bonnet gravity there is no exact holography, and therefore no
correct way to describe the physics of a dual field theory at the time
the brane crosses the black hole horizon. The Cardy-Verlinde formula
(\ref{CVGB}) will only be valid near the boundary of AdS.

\section{Discussion} \label{sect:conc}

In this paper we have attempted to extend the ideas of braneworld
holography in Einstein gravity, to Gauss-Bonnet gravity. We have found
that there exists a holographic description of a critical brane moving
in a Gauss-Bonnet AdS black hole bulk, but only when it is close to
the boundary. This is in contrast to Einstein gravity, when a
holographic description can be found even when the brane is not near
the AdS boundary. This has important implications when one considers
the Cardy-Verlinde formul{\ae} for the dual field theories on the
brane. It was previously thought that one could not cast the
thermodynamic quantities for the CFT dual to a Gauss-Bonnet AdS bulk
into a Cardy-Verlinde like formula. However, by making approximations
consistent with the limit in which the holographic description is
valid, it turns out that one can.

Finding a Cardy-Verlinde formula for the CFT with the Gauss-Bonnet AdS
dual enabled us to compare its properties with its analogue in
Einstein gravity. In particular, if we evaluate the Cardy-Verlinde
formula for Einstein gravity at the point at which the brane crosses
the black hole horizon, it gives us the Friedmann equation. This
relationship between the Cardy-Verlinde formula and the Friedmann
equation has been somewhat of a mystery, although our study of
Gauss-Bonnet braneworld holography has enabled us to shed some light
on the problem. We found that the relationship does {\it not} exist
for Gauss-Bonnet gravity.  We believe that this is because there is
no {\it exact} holography for Gauss-Bonnet gravity, and hence we
cannot make sense of the CFT physics at the time the brane crosses the
horizon -- this is explained in detail in section \ref{sect:CV}.

We would like to finish off by commenting on earlier studies of
Gauss-Bonnet braneworld
holography~\cite{Nojiri:gb1,Cho:gb2,Lidsey:gb1,Nojiri:gb2}. From
these, one draws negative conclusions about the existence of a
holographic description.  However, these studies all use an
alternative Friedmann equation, derived
in~\cite{Nojiri:hd,Nojiri:gb1}. The difference occurs because they
have different boundary terms in the Gauss-Bonnet
action~\cite{Nojiri:finite,Cvetic:boundary}. In this paper, we have
used the boundary terms derived by Myers~\cite{Myers:ghterm}. It is
encouraging that we have succeeded in gaining some positive results
using this method.  The study of braneworld cosmology using the
Friedmann equations discussed in this article has recently
begun~\cite{Lidsey:gbcos1,Lidsey:gbcos2}.  It would be an interesting
avenue of research to investigate the possible connections between the
existence of a holographic description of braneworld cosmology, and
the nature of the alternative boundary terms employed in the study of
braneworld cosmology in a Gauss-Bonnet bulk.  In particular, the
treatment of the brane as a {\it thin wall} in the bulk spacetime is
not a trivial matter in the Gauss-Bonnet braneworld
model~\cite{Deruelle:walls}.  Before one takes a thin wall limit to
approach the standard Randall-Sundrum braneworld model, the thick wall
in these models has an internal structure -- it remains an open
question as to whether the nature of this internal structure affects
the existence of a holographic description of braneworld cosmology.

\vskip .5in

\centerline{\bf Acknowledgements}
\medskip

We would like to thank Dominic Breacher, Christos Charmousis, Stephen
Davis, Shin'ichi Nojiri and Simon Ross for helpful correspondence.
JPG would particularly like to thank Ulf Danielsson for stimulating
discussions. AP would also like to thank Syksy R\"as\"anen and John
March-Russell for helpful conversations. AP was funded by PPARC.  JPG
and AP acknowledge the invaluable support of B. Bird.

\appendix
\section{Requirements for large $a$ limit} \label{app}
In this section, we will justify some of the claims made in section
\ref{sect:holography}   regarding the limit $a \gg a_H$. 

Note that the condition $h(a_H)=0$ implies that
\be \label{mu}
\tilde \al \mu=A a_H^{n-1}
\ee
where
\be
A=\tilde \al k_n^2 +\frac{\tilde \al}{a_H^2}+\left(\frac{\tilde
\al}{a_H^2}\right)^2
\ee
is of order one.

If we {\it assume} that $a_H \ll a$, we see that $4 \tilde \al
\mu/a^{n-1} \ll 1$, so that the Friedmann equation takes the form
given in equation (\ref{jamesisgay}):
\be \label{FRWapp}
H^2=\mathcal{A}-\frac{1}{a^2}+\frac{B}{A}\frac{ \mu}{a^{n-1}}
\ee
where $B$ is also of order one. For the de Sitter brane
($\mathcal{A}>0$) there are clearly a number of scenarios in which $H$
never vanishes and the brane can reach to arbitrarily large values of
$a$ (see section 5.5 in \cite{Padilla:thesis}). However, for critical
branes ($\mathcal{A}=0$) and anti-de Sitter branes ($\mathcal{A}<0$),
$a$ will have a maximum value $a_\textrm{max}$, where $H$ vanishes.

Consider the critical brane at $a_\textrm{max}$. It follows from
(\ref{FRWapp}) that
\be
\mu=\frac{A}{B}a_\textrm{max}^{n-3}
\ee
Combining this with equation (\ref{mu}), we find
\be
\frac{\tilde \al}{a_H^2}=B \left(\frac{a_H}{a_\textrm{max}}
\right)^{n-3} \ll 1
\ee
where we have used the fact that $B$ is order one.

Now consider the anti-de Sitter brane at $a_\textrm{max}$. This time
we get
\be
\mu =\frac{A}{B} \left(|\mathcal{A}|a_\textrm{max}^{n-1}+ a_\textrm{max}^{n-3}
\right)
\ee
Again, we combine this with equation (\ref{mu}) to give
\be
B=\tilde \al |\mathcal{A}|\left(\frac{a_\textrm{max}}{a_H}
\right)^{n-1}+ \frac{\tilde \al}{a_H^2} \left(\frac{a_\textrm{max}}{a_H}
\right)^{n-3}.
\ee
Since $B$ is order one, and $a_\textrm{max} \gg a_H$, we must have
\be
\tilde \al |\mathcal{A}| \ll 1, \qquad \frac{\tilde \al}{a_H^2} \ll 1.
\ee
This means that the anti-de Sitter brane is ruled out in the large $a$
analysis.

\bibliographystyle{utphys}

\bibliography{gbholography}

\end{document}